\documentclass[prl]{revtex4}

\usepackage{graphicx}
\usepackage{dcolumn}
\usepackage{bm}

\begin{document}
\renewcommand{\theequation}{\arabic{equation}}

\title{Fermion-antifermion mixing in gravitational fields\\}

\author{Giorgio Papini}
\altaffiliation[Electronic address:]{papini@uregina.ca}
\affiliation{Department of Physics
and Prairie Particle Physics
Institute, University of Regina, Regina, Sask, S4S 0A2, Canada}
\affiliation{International Institute for Advanced Scientific
Studies, 89019 Vietri sul Mare (SA),
 Italy.}

\date{\today}

\begin{abstract}
Mixing of fermion and antifermion states occurs in gravitational
interactions, leading to non-conservation of fermion number above
temperatures determined by the particle masses. We study the
evolution of a $f\,,\bar{f}$ system and calculate the cross
sections for the reactions $f\rightleftharpoons\bar{f}$. Their
values are identical in both directions. However, if $\bar{f}$
changes quickly into a lighter antiparticle, then the reaction
symmetry is broken, \emph{resulting in an increased production of matter
over antimatter}.
\end{abstract}

\pacs{PACS No.: 04.62.+v, 11.30.Fs, 95.30.Sf} \maketitle

\setcounter{equation}{0}
Of the fermions that exist as free particles, baryons appear to satisfy,
with good accuracy, a number conservation law in the strong, electromagnetic
and weak interactions. The same holds true, separately, for the numbers of
the lepton families. It is however thought that baryon and lepton numbers are
not exactly conserved quantities. This is sometimes invoked as a possible
way to explain the value $n_{B}/n_{\gamma}\sim 10^{-9}$ for the baryon to photon ratio observed in the
universe and which is at the heart of the baryosynthesis problem \cite{kolb}.

It is shown below that fermion numbers need not be conserved in gravitational interactions.

We consider external gravitational fields for which solutions
of the covariant Dirac equation \cite{caip,dinesh,lamp}
\begin{equation}\label{CDE}
  [i\gamma^{\mu}(x){\cal D}_\mu-m]\Psi(x)=0\,,
\end{equation}
and of other covariant wave equations \cite{pap,pap3,pap1} can be
constructed. These solutions are exact to first order in the
metric deviation $\gamma_{\mu\nu}=g_{\mu\nu}- \eta_{\mu\nu}$,
where $\eta_{\mu\nu}$ is the Minkowski metric and can be
calculated explicitly by means of path integrals if the solution
of the free wave equation (corresponding to the same particle of
mass $m$) and $\gamma_{\mu\nu}(x)$ are known. The notations are
those of \cite{lamp,rad}, in particular
$\mathcal{D}_{\mu}=\nabla_{\mu}+i\Gamma_{\mu}(x)$, $\nabla_{\mu}$
is the covariant derivative and $\Gamma_{\mu}(x)$ the spin
connection. A semicolon and a comma are frequently used to
indicate covariant and partial derivatives respectively. The first
order solutions of (\ref{CDE}) are of the form
\begin{equation}\label{E}
  \Psi(x) = {\hat T}(x) \psi(x)\,,
\end{equation}
where $\psi(x)$ satisfies the usual flat spacetime Dirac equation
\begin{equation}\label{DE}
\left(i\gamma^{\hat{\mu}}\partial_{\mu}-m\right)\psi(x)=0\,,
\end{equation}
and is represented here by the positive (particle) and negative (antiparticle) energy solutions
$\psi(x)=u(\vec{k})e^{-ik_{\mu}x^{\mu}}$ and $\psi^{(1)}(x)=v(\vec{k})e^{ik_{\mu}x^{\mu}}$ respectively,
where the spin-up ($\uparrow$) and spin-down ($\downarrow$) components of the
spinors $u$ and $v$ obey the well-known relations \cite{jauch}
\begin{equation}\label{spin}
u_{\downarrow}=\gamma^{5}v_{\uparrow}\,,\qquad v_{\downarrow}=\gamma^{5}u_{\uparrow}\,.
\end{equation}
The matrices $\gamma^{\hat{\mu}}$ in (\ref{DE}) are the usual
constant Dirac matrices. The operator $\hat{T}$ is given by
\cite{lamp}
\begin{equation}\label{T}
    \hat{T}=
  -\frac{1}{2m}\left(-i\gamma^{\mu}(x)\mathcal{D}_{\mu}-m\right)e^{-i\Phi_{T}}\,,
\end{equation}
where
\begin{equation}\label{PHIS}
\Phi_{T}=\Phi_{S}+\Phi_{G}\,,\qquad
\Phi_{S}(x)=\int_{P}^{x}dz^{\lambda}\Gamma_{\lambda}(z)\,,
\end{equation}
and
 \begin{equation}\label{PH}
  \Phi_{G}(x)=-\frac{1}{4}\int_P^xdz^\lambda\left[\gamma_{\alpha\lambda,
  \beta}(z)-\gamma_{\beta\lambda, \alpha}(z)\right]\left[\left(x^{\alpha}-
  z^{\alpha}\right)k^{\beta}-\left(x^{\beta}-z^{\beta}\right)k^{\alpha}\right]+
  \frac{1}{2}\int_P^xdz^\lambda\gamma_{\alpha\lambda}(z)k^{\alpha}\,.
\end{equation}
In (\ref{PHIS}) and (\ref{PH}), the path integrals are taken along
the classical world line of the particle starting from a point
$P$ and $\Phi_{G}\sim \mathcal{O}(\gamma_{\mu\nu})$ is the covariant Berry phase \cite{caip1}.
It follows from (\ref{spin}), $\gamma^{5} \equiv i\gamma^{\hat{0}}\gamma^{\hat{1}}
\gamma^{\hat{2}}\gamma^{\hat{3}}\,,\left\{\gamma^{5},\gamma^{\hat{\mu}}\right\}
=0$, $\sigma^{{\hat \alpha}{\hat \beta}}=\frac{i}{2}[\gamma^{\hat
\alpha}, \gamma^{\hat \beta}]$ and
\begin{equation}\label{ga}
  \gamma^\mu(x)=e^\mu_{\hat \alpha}(x) \gamma^{\hat
  \alpha}\,,\qquad
  \Gamma_\mu(x)=-\frac{1}{4} \sigma^{{\hat \alpha}{\hat \beta}} e^\nu_{\hat \alpha}e_{\nu\hat{\beta};\, \mu}\,,\qquad
    [\gamma^{5},\Gamma^{\mu}]=0\,,
\end{equation}
that if $\Psi(x)=e^{-ik_{\mu}x^{\mu}}\hat{T}u$ is a solution of (\ref{CDE}), then
$\Psi^{(1)}(x)=e^{ik_{\mu}x^{\mu}}\hat{T}_{1}v$
also is a solution of (\ref{CDE}), and $\hat{T}_{1}=\gamma^{5}\hat{T}\gamma^{5}$.
It is useful
to further isolate the gravitational contribution in the vierbein
fields by writing
$e^{\mu}_{\hat{\alpha}}\simeq\delta^{\mu}_{\hat{\alpha}}+h^{\mu}_{\hat{\alpha}}$.
We therefore obtain
\begin{equation}\label{T}
{\hat
T}=\hat{T}_{0}+\hat{T}_{G}=\frac{1}{2m}\left\{\left(1-i\Phi_{G}\right)\left(m+\gamma^{\hat{\alpha}}
k_{\alpha}\right)-i
\left(m+\gamma^{\hat{\alpha}}k_{\alpha}\right)\Phi_{S}+
  \left(k_{\beta}h^{\beta}_{\hat{\alpha}}+\Phi_{G,\alpha}\right)\gamma^{\hat{\alpha}}\right\}\,,
\end{equation}
where
$\hat{T}_{0}=\frac{1}{2m}\left(m+\gamma^{\hat{\alpha}}k_{\alpha}\right)$
and $\hat{T}_{G}$ contains the (first order) gravitational
corrections. Similarly, we find
\begin{equation}\label{NMS}
{\hat T}_1=\hat{T}_{10}+\hat{T}_{1G}=\frac{1}{2m}\left\{\left(1-i\Phi_{G}\right)\left(m-
\gamma^{\hat{\alpha}}k_{\alpha}\right)-i\left(m-\gamma^{\hat{\alpha}}k_{\alpha}\right)\Phi_{S}-
  \left(k_{\beta}h^{\beta}_{\hat{\alpha}}+\Phi_{G,\alpha}\right)\gamma^{\hat{\alpha}}\right\}\,,
\end{equation}
where $\hat{T}_{10}=\frac{1}{2m}\left(m-\gamma^{\hat{\alpha}}k_{\alpha}\right)$.
Notice that $\psi(x)$ and $\psi^{(1)}(x)$ as well as $\Psi(x)$ and $\Psi^{(1)}(x)$
describe particles of the same mass $m$ as required by $CPT$-invariance.

In general the eigenstates $U^{\pm}=
1/\sqrt{2}(u \pm v)$ of $\gamma^{5}$ and the eigenstates
$u$  and $v$ of $\hat{T}_{0}$ are not the same
and $\hat{T},\,\hat{T}_{1}$ mix $u$  and $v$,
thus producing $f\bar{f}$ transitions. A similar mixing phenomenon
occurs in the $K^{0},\,\bar{K}^{0}$ system where the $CP$
eigenstates are mixed by the weak interaction Hamiltonian
resulting in $CP$ violation. It is the fermion number conservation
that is violated in our work and the mixing is carried out by
the gravitational contributions $\hat{T}_{G}$ and $\hat{T}_{1G}$. This can be seen as follows.

The time evolution of a $f,\bar{f}$ system in a gravitational field can be
written in the form
\begin{equation}\label{PS}
 |\Phi(t) \rangle = \alpha_{0} |\Psi(t)\rangle
 +\beta_{0} |\Psi^{(1)}(t)\rangle=\alpha_{0}\hat{T}(t)|\psi(t)\rangle+\beta_{0}\hat{T}_{1}(t)|\psi^{(1)}(t)\rangle
 \equiv\alpha(t)|\psi(t)\rangle+\beta(t)|\psi^{(1)}(t)\rangle\,,
\end{equation}
where $|\alpha_{0}|^2+|\beta_{0}|^2=1$, from which we obtain
\begin{equation}\label{al}
\alpha(t)= \langle \psi|\Phi(t)\rangle
=\alpha_{0}\langle\psi|\hat{T}|\psi\rangle +\beta_{0}\langle
\psi|\hat{T}_{1}|\psi^{(1)}\rangle\,;\beta(t)=\langle
\psi^{(1)}|\Phi(t)\rangle =\alpha_{0}\langle
\psi^{(1)}|\hat{T}|\psi \rangle +\beta_{0}\langle
\psi^{(1)}|\hat{T}_{1}|\psi^{(1)}\rangle \,. \qquad
\end{equation}
If at $t=0$ the gravitational field is not present, then
$\hat{T}_{G}=0\,,\hat{T}_{1G}=0$ and
$\alpha_{0}=\alpha(0)\,,\beta_{0}=\beta(0)$. It follows from
(\ref{al}) that as the $f\,,\bar{f}$ system propagates in a
gravitational field, transitions $f\rightleftharpoons\bar{f}$ can
take place.

In order to obtain the transition probabilities
$|\alpha(t)|^{2}\,,|\beta(t)|^{2} $ from (\ref{al}) in a concrete
case, we choose for simplicity
\begin{equation}\label{psi0}
\psi(x)=f_{0,R}e^{-ik_\alpha x^\alpha}=\sqrt{\frac{E+m}{2m}}
  \left(\begin{array}{c}
                f_{R} \\
                 \frac{\sigma^{3} k}{E+m}\, f_{R} \end{array}\right)
                 \,e^{-ik_\alpha x^\alpha}\,,
\end{equation}
where $ f_{R}$ is the positive helicity eigenvector.
The normalizations are $\langle \psi|\psi\rangle =1$, where $\langle
\psi|= \langle\psi^{\dag}|\gamma^{\hat{0}}$, $\langle
\psi^{(1)}|\psi^{(1)}\rangle =-1$ and $\langle
\psi|\psi^{(1)}\rangle = \langle \psi^{(1)}|\psi\rangle =0$. In
addition, we need explicit expressions of the metric components
for the purpose of calculating $\hat{T}$ and $\hat{T}_{1}$. We
choose the Lense-Thirring metric \cite{lense}, represented, in its
post-Newtonian form, by
\begin{equation}\label{LTmetric}
 \gamma_{00}=2\phi\,, \quad \gamma_{ij}=2\phi\delta_{ij}\,, \quad
\gamma_{0i}=h_i=\frac{2}{r^3}(\textbf{J}\wedge\textbf{r})_i\,,
 \end{equation}
where
\begin{equation}\label{LTmetric1}
  \phi=-\frac{GM}{r}\,, \quad
  \textbf{h}=\frac{4GMR^{2}\omega}{5r^3}(y,-x,0)\,,
\end{equation}
and  \textit{M}, \textit{R}, $ \mathbf{\omega}=(0,0,\omega)$ and $
\mathbf{J}$ are mass, radius, angular velocity and angular
momentum of the source. This metric is non trivial, has no Newtonian
counterpart and describes the physically significant case of a rotating source. By using the
freedom allowed by local Lorentz
transformations, the vierbein field to
$\mathcal{O}(\gamma_{\mu\nu})$ is
\begin{equation}\label{3.5}
 e^0_{\hat{i}}=0\,{,}\quad
 e^0_{\hat{0}}=1-\phi\,{,}\quad
 e^i_{\hat{0}}=h_i\,{,}\quad
 e^l_{\hat{k}}=\left(1+\phi\right)\delta^l_k\,.
 \end{equation}
 Without loss of generality, we consider particles starting from
 $z=-\infty$, and propagating, with impact parameter $b\geq R$,
 along $x=b\,,y=0$ in the field of the rotating source. We also set $k^{3}\equiv k$ and $k^{0}\equiv E$.

We can now return to (\ref{al}). If originally the system is an
antifermion, then $ \alpha_{0}=0\,, \beta_{0}=1$, $
|\Phi(t)\rangle = \hat{T}_{1} |\psi^{(1)}\rangle $ and from
(\ref{PS}) we obtain
\begin{equation}\label{al2}
\alpha(t)= \langle \psi|\hat{T}_{1}|\psi^{(1)}\rangle= \frac{e^{iq_{\alpha}x^{\alpha}}}{2m}\left\{-\langle \psi|
\left[Eh_{\hat{0}}^{0}\gamma^{\hat{0}}+\left(-kh_{\hat{3}}^{3}+\left(\frac{E^2}{k}+k\right)
\phi(z)\right)\gamma^{\hat{3}}\right]|\psi^{(1)}\rangle\right\}\,,
\end{equation}
where $q_{i}=k'_{i}+k_{i}$, $k'_{i}$ refers to $\langle \psi|$ and $q_{0}=2E$ because the external field is time-independent.
The probability of the transition $\psi^{(1)} \rightarrow \psi$,
which, because of the coupling of spin to gravity, violates $\bar{f}$ number conservation,
follows from (\ref{al2}) and is
\begin{equation}\label{BL}
P_{\psi^{(1)}\rightarrow \psi}=|\alpha(t)|^{2} =\left[\frac{1}{2m^2} (k^2 - \frac{E^3}{k})\right]^2 \phi^2(z) \,.
\end{equation}
In the approximation  $k\gg m$ we obtain from (\ref{BL}) $P_{\psi^{(1)}\rightarrow \psi}=
(3/4)^2 \phi^2$, while  $P_{\psi^{(1)}\rightarrow \psi}=(m/2k)^2 \phi^2$ for $k\ll m$.
The latter result leads in the limit $k\rightarrow 0$ to a well known infrared divergence
that will be discussed below.

If we choose $ \alpha_{0}=1, \beta_{0}=0$, then $|\beta(t)|^2$
represents the probability for the inverse process
$\psi\rightarrow \psi^{(1)}$. We find
\begin{equation} \label{BE}
P_{\psi\rightarrow \psi^{(1)}}=|\beta(t)|^2=|\langle \psi^{(1)}|\hat{T}|\psi\rangle|^{2}=
|\langle \psi^{(1)}|\gamma^{5}\hat{T_{1}}\gamma^{5}|\psi\rangle|^{2}=
|\langle \psi|\hat{T}_{1}|\psi^{(1)}\rangle|^{2}=P_{\psi^{(1)}\rightarrow \psi}\,.
\end{equation}
According to (\ref{BL}) and (\ref{BE}), the transitions proceed in both directions with
the same probability, with no contribution from the source's rotation when $\vec{k}$ is
in the $z$-direction, as specified.

With the help of (\ref{BL}) we can also calculate the cross section for the
$\bar{f}\rightarrow f$ process which corresponds to a change $|\Delta L|=2$
for leptons and $|\Delta B| =2$ for baryons. We get \cite{renton}
\begin{equation}\label{CS1}
\frac{d\sigma}{d\Omega}=\frac{|\langle\psi|\hat{T}_{1}|\psi^{(1)}\rangle|^{2}}
{E k}k_{f}^{2}\delta(E_{f}-E)dk_{f}=|\langle\psi|\hat{T}_{1}|\psi^{(1)}\rangle|^{2}\,,
\end{equation}
where use has been made of the relations $k_{f}dk_{f}=E_{f}dE_{f}\,, k_{f}^{2}=E_{f}^{2}-m^2=
E^2 -m^2 =k^2$, integration over $E_{f}$ has been performed and the index $f$
indicates momentum and energy in the final state. There is, of course, no energy
transfer to the external field, since the field is static.
We also find from (\ref{BL})
\begin{equation}\label{TRA2}
|\langle\psi|\hat{T_{1}}|\psi^{(1)}\rangle|^2 = \left[\frac{1}{2m^2} (k^2 - \frac{E^3}{k})
\right]^2\phi(\tilde{k})^2\,,
\end{equation}
where $\phi(\tilde{k})=-(GM/\pi)K_{0}(b\tilde{k})$ is the Fourier transform
of $\phi(z)=-GM/\sqrt{z^2 +b^2}$, the momentum transfer is $\tilde{k}=|\vec{k}_{f}-
\vec{k}|=2k\sin(\theta/2)$ and $\theta$ is the angle that $\vec{k}_{f}$ makes with the $z$-axis.
Integrating over the angles, we obtain from (\ref{CS1}) the total cross-section
\begin{equation}\label{CST}
\sigma = \frac{1}{\pi}\left(\frac{k^3 -E^3}{m^{2}k}\right)^2
\left(GM\right)^2
\left[\frac{1}{4b^2k^2}+K^2_{0}(2bk)-K^2_{1}(2bk)\right]\,.
\end{equation}
On account of the exponential decay of the Bessel functions,
(\ref{CST}) can be approximated by
\begin{equation}\label{CSA}
\sigma \approx \frac{1}{4\pi}\left[\gamma\left(\beta-\frac{1}{\beta^2}\right)\right]^2\left(\frac{GM}{bm}\right)^{2}\,,
 \end{equation}
for all physical values of $bk$. In (\ref{CSA}), $\beta$ and
$\gamma$ are the usual Lorentz factors and $k=m\beta\gamma$. It
also follows from (\ref{TRA2}) and $\hat{T}=\gamma^{5}
\hat{T_{1}}\gamma^{5}$ that the transition amplitude
$\langle\psi^{(1)}|\hat{T}|\psi\rangle$ leads to the same
cross-section (\ref{CST}). From $0\leq\beta \leq 1$, one obtains
$T\geq m/\kappa\equiv T_{c}$ and also $T_{c}\simeq
1.2\cdot10^{4}K$ for neutrinos of mass $m\sim 1 eV$ and $T_{c}
\simeq 1.1 \cdot10^{13}K$ for nucleons. Expressing $\beta$ and
$\gamma$ as temperatures, we can re-write (\ref{CSA}) in the form
\begin{equation}\label{CSB}
\sigma = \frac{1}{4\pi}(\frac{GM}{bm})^{2}\left(\sqrt{\tilde{T}^2 -1}-\frac{\tilde{T}^3}{\tilde{T}^2 -1}\right)^2\,,
\end{equation}
where $\tilde{T}\equiv T/T_{c}$. The transitions discussed are forbidden for $T<T_{c}$,
have vanishing value for $\tilde{T}\rightarrow \infty$ and diverge for
$\tilde{T}\rightarrow 1$ as expected. This divergence, already met in
connection with (\ref{BL}), arises as a consequence of the implied requirement
that particles interact without the emission of gravitons (or photons in the electromagnetic case)
\cite{jauch}). This requirement can not in fact be realized physically.
Processes in which gravitons with energy less than the energy resolution
$\Delta\epsilon$ of the apparatus
are in fact indistinguishable from those in which gravitons of energy less or
equal to $\Delta\epsilon$ are also emitted and reabsorbed by the particles. The natural cutoff
for the graviton energy is therefore $\Delta\epsilon$. It is also shown in the literature
that when the gravitational \cite{weinberg} and electromagnetic \cite{jauch} fields are quantized,
all infrared divergencies disappear.

\emph{Summary and discussion}. It is possible for baryons and
leptons to achieve the required energies in astrophysical
situations. If, for the sake of numerical comparisons, we take
$GM/b\sim 0.1$, which applies in the vicinity of a black hole, and
$\beta\sim 0.95$, then the cross section value is $\sigma \sim 8
\cdot 10^{-14}cm^2$ for $1 eV$ neutrinos, while we find
$\sigma\sim 8 \cdot 10^{-32}cm^{2}$ for baryons of mass $m \sim 1
GeV$. It therefore follows from (\ref{CSA}) and (\ref{CSB}) that,
for the same value of $\beta$, the baryon number is more likely to
be conserved ($\sigma$ smaller) than the neutrino number, i.e. the
$ B$-violating processes are suppressed relative to those
violating $ L$. The actual value of $\sigma$ depends on
$(GM/b)^2$. It may therefore be useful to re-examine attentively
some of the processes that take place close to compact, massive
objects.

Gravity behaves similarly on the cosmological scale and does not
conserve the fermion number at lower momenta and higher
temperatures. The values of $T_{c}$ given above indicate that $B$
conservation occurred in the universe at the time of primordial
nucleosynthesis, while $L$ conservation is relatively recent, thus
resulting in the coexistence of two fermion populations at
temperatures $T_{c,L}\leq T\leq T_{c,B}$, with the leptons
agitated by the $f \rightleftharpoons \bar{f}$ transitions, and
the baryons dormant (neglecting, of course, the effect of any
other concomitant interactions).

The cross section for $f \rightleftharpoons\bar{f}$ reactions is
the same in both directions, hence the $ B$- and $ L$-violating
processes do not produce more fermions than antifermions. The
baryosynthesis problem can not therefore be solved using the
gravitational non-conservation of fermion number without
additional assumptions regarding the initial distributions of
matter and antimatter, or the breakdown of thermal equilibrium.

There is, however, an important, additional possibility suggested
by (\ref{CSA}). If, in fact, $\bar{f}$ of mass $m$ produced in the
reaction $f\rightarrow \bar{f}$ changes rapidly into an
antifermion of mass $m'< m$, then the cross section for the
inverse $\bar{f}'\rightarrow f'$ process can no longer
re-establish symmetry between matter and antimatter. This reaction
cross section does in fact increase, relative to the original one,
by a factor $\sigma'/\sigma\sim
(m/m')^{2}[\gamma'(\beta'-1/\beta'^{2})]^{2}/[\gamma(\beta-1/\beta^{2})]^{2}$
which can be large, depending on $(m/m')^{2}$ and the reaction
kinematics. \emph{The result is therefore an increase in the production
of matter over antimatter}. Any further analysis of the mechanism's relevance to 
baryon asymmetry depends on $ m$ and $m'$ and their experimental signatures and is not pursued here.

Scalar-tensor theories of gravitation based on early efforts by a number of authors
\cite{Fierz}, \cite{Jordan}, \cite{BD} have been recently considered for
the purpose of obtaining a varying gravitational constant that could in
principle enhance the strength of gravity \cite{CE},\cite{RW},\cite{KR}. Would the mixing
mechanism discussed above be affected by the introduction of a
running gravitational constant and concomitant scalar field? Unfortunately,
no useful, first order tensor-scalar solutions of (\ref{CDE}) exist at present.
Nonetheless, a qualitative indication of the role
played by a scalar field and running gravitational constant can be obtained from
(\ref{CDE}) by means of the conformal transformation $ g_{\mu\nu}=G\Lambda(x)\eta_{\mu\nu}$
which separates the scalar from the tensor component of the theory.
In this simple case, (\ref{CDE}) can be solved \cite{LP}, the operators
$\hat{T}$ and $\hat{T}_1$ given by (\ref{T}) and (\ref{NMS}) become
\begin{eqnarray}\label{TC}
\hat{T}^{c}=\frac{1}{2m}\left\{\left(m + \gamma^{\hat{\alpha}}k_{\alpha}\right)+\frac{3i}{4}\left(\ln G\Lambda(x)\right)_{,\alpha}\gamma^{\hat{\alpha}}\right\}\\
\hat{T}_{1}^{c}=\frac{1}{2m}\left\{\left(m - \gamma^{\hat{\alpha}}k_{\alpha}\right)-\frac{3i}{4}\left(\ln G\Lambda(x)\right)_{,\alpha}\gamma^{\hat{\alpha}}\right\}
\end{eqnarray}
and still confirm the presence of the symmetry breaking mechanism.
However, in view of the limitations on the structure of $\Lambda(x)$
discussed in \cite{LP}, more precise conclusions must await the detailed study
of a meaningful physical model.

The introduction of back reaction terms discussed in \cite{rad},
while increasing the cross-sections over appropriate time
intervals, does not alter the symmetry of the original
distribution. The back reaction would however increase the
cross-section of the process $\bar{f}'\rightarrow f'$ discussed
above, and of a reaction for which $CP$ violation has been
observed, as in the time evolution of a $K^{0}\bar{K}^{0}$ system,
if the latter process occurred in the neighborhood of a
gravitational source of appropriate strength. The model of back
reaction discussed in \cite{rad} is based in fact on the
introduction of a dissipation term in the wave equation of the
particle propagating in an external gravitational field. It is
then shown that a gravitational perturbation can grow rapidly, as
with a fluid heated from below in which a small disturbance in the
wave equation grows rapidly as soon as convection starts. The
covariant wave equation of a kaon in the external gravitational
field approximation has the solution $\varphi(x) =
e^{-i\Phi_{G}(x)}\varphi_{0}(x)$ where $\varphi_{0}(x)$ satisfies
the equation \cite{pap3}
\begin{equation}\label{flat}
\left(\eta^{\mu\nu}\partial_{\mu}\partial_{\nu}+ m^2\right)\varphi_{0}(x)=0\,.
\end{equation}
 It follows \cite{rad} that the addition to (\ref{flat}) of a dissipation term $ -2m \xi
 \partial_{0}\varphi_{0}$
 where $\xi = \varepsilon (\frac{m}{k_{0}}\frac{GM}{2b})^2$ and $\varepsilon$ is
 a dimensionless, arbitrary parameter $0\leq \varepsilon\leq 1$, transforms the
 solution of the covariant wave equation into
 $\varphi(x) = e^{-i\Phi_{G}(x)}e^{m\xi x_{0}}\varphi_{0}(x)$ and the overlap of the
 kaon eigenstates $K_{L}$ and $K_{s}$, which is a measure of $CP$ violation, into $\langle K_{L}|K_{s}\rangle_{G}=
 \langle K_{L}|K_{s}\rangle e^{2m\xi x_{0}}$. Then the increase in $CP$ violation
 could become large even if the growth of the exponential term were restricted by competing effects \cite{rad}.


\begin{thebibliography}{99}
\bibitem{kolb} See, e. g., P. D. B. Collins, A. D. Martin, E. J. Squires, {\it Particle Physics and Cosmology} (John Wiley and Sons, New York, 1989).
\bibitem{caip} Phys. Rev. Lett. {\bf 66}, 1259 (1991).
\bibitem{dinesh} D. Singh, G. Papini, Nuovo Cim. B {\bf 115}, 223 (2000).
\bibitem{lamp} G. Lambiase, G. Papini, R. Punzi, G. Scarpetta, Phys. Rev. D {\bf 71}, 073011 (2005).
\bibitem{pap} G. Papini, Phys. Rev. D {\bf 75}, 044022 (2007).
\bibitem{pap3} G. Papini, Gen. Rel. Gravit. {\bf 40}, 1117 (2008).
\bibitem{pap1} G. Papini, G. Scarpetta, A. Feoli, G. Lambiase, Int. J. Mod. Phys. D {\bf 18}, 485 (2009).
\bibitem{rad} G. Papini, Phys. Rev. D {\bf 82}, 024041 (2010).
\bibitem{jauch} J. M. Jauch and F. Rohrlich, \emph{The Theory of Photons and Electrons} (Springer, New York 1976).
\bibitem{caip1} Y. Q. Cai, G. Papini, Mod. Phys. Lett. A {\bf 4}, 1143 (1989); Class. Quantum Grav. {\bf 7}, 269 (1990).
\bibitem{lense} J. Lense and H. Thirring, Z. Phys. {\bf 19}, 156(1918);
(English translation: B. Mashhoon, F.W. Hehl and D.S. Theiss, Gen. Rel. Grav. {\bf 16}, 711(1984)).
\bibitem{renton} See, e. g., Peter Renton, \emph{ Electroweak Interactions} (Cambridge University Press, New York, 1990).
\bibitem{weinberg} Steven Weinberg, Phys. Rev. {\bf 140}, B516 (1965).
\bibitem{Fierz} M. Fierz, Helv. Phys. Acta {\bf 29}, 128 (1956).
\bibitem{Jordan} P. Jordan, Z. Phys. {\bf 157}, 112 (1959).
\bibitem{BD} C. Brans, R. H. Dicke, Phys. Rev. {\bf 124}, 925 (1961).
\bibitem{RW} M. Reuter, H. Weyer, J. Cosmol. Astropart. Phys. 12, 001 (2004).
\bibitem{KR} B. Koch, I. Ramirez, Classical Quantum Gravity {\bf 28}, 055008 (2011).
\bibitem{CE} Yi-Fu Cai, Damien A. Easson, Phys. Rev. D {\bf 84}, 103502 (2011).
\bibitem{LP} G. Lambiase, G. Papini, Nuovo Cim. B {\bf 113}, 1047 (1998).
\end{thebibliography}
\end{document}